\documentclass[sigconf]{acmart}

\AtBeginDocument{%
  }

\setcopyright{none}

\acmConference[preprint]{}{}{}

\acmISBN{978-1-4503-XXXX-X/18/06}

\usepackage{spverbatim}
\usepackage{enumitem}

\usepackage{color}
\usepackage{framed}
\usepackage{tcolorbox}
\usepackage{tikz}
\usepackage{listings}

\newtcolorbox{myverbbox}{
  colback=gray!10,
  colframe=gray!80!black,
  width=\linewidth,
  boxrule=0pt,
  left=2mm,
  right=2mm,
  top=2mm,
  bottom=2mm,
  before=\noindent,
  after=\par,
  boxsep=0pt, 
  coltitle=black,
  fonttitle=\ttfamily,
  fontupper=\ttfamily, 
}
\usepackage[whole]{bxcjkjatype}
\begin{document}

\title{MagicItem: Dynamic Behavior Design of Virtual Objects with Large Language Models in a Consumer Metaverse Platform}

\author{Ryutaro Kurai}
\affiliation{%
  \institution{Cluster, Inc.}
  \city{Tokyo}
  \country{Japan}
}
\affiliation{%
  \institution{Nara Institute of Science and Technology}
  \city{Nara}
  \country{Japan}
}
\email{r.kurai@cluster.mu}

\author{Takefumi Hiraki}
\affiliation{%
  \institution{Cluster Metaverse Lab}
  \city{Tokyo}
  \country{Japan}
}
\email{t.hiraki@cluster.mu}

\author{Yuichi Hiroi}
\affiliation{%
  \institution{Cluster Metaverse Lab}
  \city{Tokyo}
  \country{Japan}
}
\email{y.hiroi@cluster.mu}

\author{Yutaro Hirao}
\affiliation{%
  \institution{Nara Institute of Science and Technology}
  \city{Nara}
  \country{Japan}
}
\email{yutaro.hirao@is.naist.jp}

\author{Monica Perusquia-Hernandez}
\affiliation{%
  \institution{Nara Institute of Science and Technology}
  \city{Nara}
  \country{Japan}
}
\email{m.perusquia@is.naist.jp}

\author{Hideaki Uchiyama}
\affiliation{%
  \institution{Nara Institute of Science and Technology}
  \city{Nara}
  \country{Japan}
}
\email{hideaki.uchiyama@is.naist.jp}

\author{Kiyoshi Kiyokawa}
\affiliation{%
  \institution{Nara Institute of Science and Technology}
  \city{Nara}
  \country{Japan}
}
\email{kiyo@is.naist.jp}
\renewcommand{\shortauthors}{Kurai et al.}


\begin{abstract}
To create rich experiences in virtual reality (VR) environments, it is essential to define the behavior of virtual objects through programming. However, programming in 3D spaces requires a wide range of background knowledge and programming skills. Although Large Language Models (LLMs) have provided programming support, they are still primarily aimed at programmers. In metaverse platforms, where many users inhabit VR spaces, most users are unfamiliar with programming, making it difficult for them to modify the behavior of objects in the VR environment easily. Existing LLM-based script generation methods for VR spaces require multiple lengthy iterations to implement the desired behaviors and are difficult to integrate into the operation of metaverse platforms.
To address this issue, we propose a tool that generates behaviors for objects in VR spaces from natural language within Cluster, a metaverse platform with a large user base. By integrating LLMs with the Cluster Script provided by this platform, we enable users with limited programming experience to define object behaviors within the platform freely. We have also integrated our tool into a commercial metaverse platform and are conducting online experiments with 63 general users of the platform. The experiments show that even users with no programming background can successfully generate behaviors for objects in VR spaces, resulting in a highly satisfying system. Our research contributes to democratizing VR content creation by enabling non-programmers to design dynamic behaviors for virtual objects in metaverse platforms.
\end{abstract}

\begin{teaserfigure}
  \includegraphics[width=\textwidth]{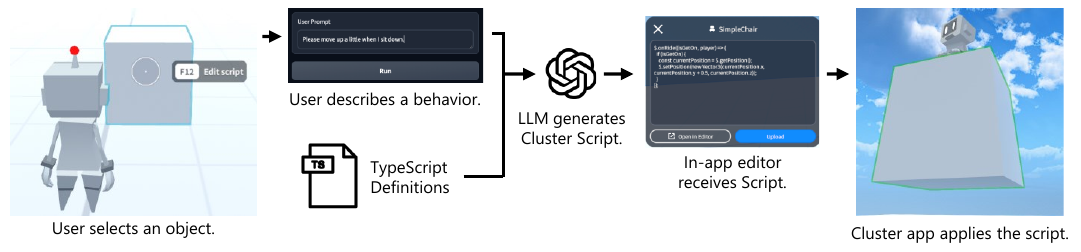}
  \caption{When the user selects an object in the Cluster client application, they describe the behavior of the chosen object into the tool we made. Consequently, the LLM generates a Cluster Script based on the description that the user wrote. Then, the script editor in the Cluster client application receives the script, and finally, the Cluster application applies the script.}
  \Description{System Pipeline}
  \label{fig:teaser}
\end{teaserfigure}

\received{19 June 2024}
\received[revised]{12 March 2009}
\received[accepted]{5 June 2009}

\maketitle

\section{Introduction}
The advancement of virtual reality (VR) technology has led to a surge in the popularity of metaverse platforms, which provide innovative ways for people to communicate, socialize, engage in economic activities, and unleash their creativity through avatars in immersive 3D virtual environments.
By 2022, more than 171 million users worldwide are expected to be using various social VR platforms such as VRChat~\cite{vrchat}, Roblox~\cite{roblox}, and Cluster~\cite{clsuter}. 

The key feature of the metaverse is the ability for users to freely create and collaborate in 3D spaces and interactions that would be difficult to realize in the physical world due to constraints.
However, programming in 3D spaces requires extensive expertise in various domains, such as 3D modeling, animation, and interaction design, and acquiring these skills requires significant time and effort.
This creates a high barrier to entry for non-programmers who wish to participate in creative activities in the metaverse.

The emergence of Large Language Models (LLMs) has dramatically advanced the field of programming assistance using natural language processing. Integrating LLMs to support 3D content creation is expected to greatly enhance users' creativity in the metaverse, especially those without prior programming experience. When integrating LLM-based scripting capabilities into metaverse platforms, it is critical to ensure that user natural language input is seamlessly translated into object behavior in real-time, especially in active applications with multiple concurrent users. However, previous studies have been limited to small-scale experiments in experimental environments and integration with commercial metaverse platforms and large-scale user evaluation has not been conducted.

To address these challenges, we propose MagicItem, a system that allows users to intuitively generate object behaviors in VR spaces using natural language on Cluster, a commercial metaverse platform.
Cluster has more than 35 million cumulative users and enables real-time description and execution of 3D object behaviors using its proprietary language, Cluster Script, an extended processing system of JavaScript.
This study uses GPT-4 to generate Cluster Script code that translates the user's natural language input into object behavior. This code is seamlessly embedded into the objects in Cluster, allowing the user to test the behavior in a 3D space.

To evaluate the effectiveness of the proposed system, we conducted the first large-scale online experiment on Cluster with 63 participants. Although nearly half of the participants had no prior experience with the Cluster script, the majority were able to modify object behavior. Usability testing included metrics of code generation subjective ratings using the NASA Task Load Index (NASA-TLX)~\cite{Nasa1986-vw} and the System Usability Scale (SUS)~\cite{Brooke1996-kw}. 
Based on these evaluation results and the participants' comments, we discuss the limitations and future research directions of our system.
The main contributions of this work are as follows:

\begin{itemize}[leftmargin=*]
\item Developing a natural language-based object behavior generation tool using LLMs on the large-scale, commercial metaverse platform Cluster.
\item Integrating LLMs with Cluster Script to enable seamless object control while ensuring synchronization between multiple users.
\item Conducting an empirical evaluation of the effectiveness and usability of the proposed system through a large-scale experiment using an online experimental environment built on Cluster.
\item Providing discussion including current limitations and future research directions.
\end{itemize}

\section{Related Work}
\subsection{Social VR Platforms}
Starting with Second Life~\cite{secondlife} in the 2000s, there are a number of commercial services that allow large numbers of people to gather and communicate in a 3D space, including VRChat~\cite{vrchat}, Neos~\cite{neos}, Resonite~\cite{resonite},
RecRoom~\cite{recroom}, and Cluster~\cite{clsuter}. 
These services are attracting attention and are referred to as social VR platforms or metaverse platforms.

The features of these services are that the space is represented in 3D, that users can create spaces, and that they can share the spaces they create with other users. It is also possible to control the behavior of objects in the space using programming languages.
Services that were previously treated as online games, such as Fortnite~\cite{fortnite}, Roblox~\cite{roblox}, and Minecraft~\cite{minecraft}, are also beginning to be recognized as communication tools beyond games.
These services also enable to share user-created spaces with other users and allow control through programming languages.

There is a lot of research on avatars and the social responses they generate on these platforms~\cite{Biocca2003, Kolesnichenko2019, Latoschik2017}.
In particular, how users use spatial relationships with each other and with objects~\cite{Hindmarsh2000} and how users use non-verbal behavior~\cite{Fabri1999, Pan2017, Yee2007} have been studied extensively.
In addition, in recent years, there has been a progressive exploration of topics such as supporting long-term relationships~\cite{Moustafa2018}, the impact of avatar expression on trust formation~\cite{Pan2017}, the requirements for preventing harassment~\cite{Blackwell2019}, and the user's own perceptions of their avatars~\cite{Freeman2020}.

Furthermore, there is also Ubiq~\cite{Friston2021-td}, a more specialized social VR platform for research.
Ubiq is an open-source platform that provides core functionality for many social VR systems, such as connection management, voice, and avatars, in an easily extensible format.
Due to its features, many studies have been proposed on social VR platforms using Ubiq~\cite{Steed2022, Numan2022, Numan2023-nx, Giunchi2024-gt}.
However, such research platforms do not have users who use them daily, and it has been difficult to conduct large-scale user experiments.

\subsection{AI-Assisted 3D Space Design and Control}
Designing and controlling 3D spaces is a very important element in VR. Making this possible efficiently with AI, or even for beginners with little technical knowledge, is recognized as an important issue. 3D authoring tools such as Blender~\cite{blender} are widely used to design 3D objects. On the other hand, the behavior of objects in 3D space is typically controlled by game engines such as Unity~\cite{unity} and Unreal Engine~\cite{unreal-engine} or platform-specific development environments. Previous research has attempted to automate the design and control of a 3D space by combining these environments with AI systems.

Ubiq-Genie~\cite{Numan2023-nx} implements sub-functions such as speech-to-text (STT), text-to-speech (TTS), and text synthesis, linked to a large language model (LLM).
Neural Canvas~\cite{Shen2024-dq} uses generative AI to create detailed 3D spaces from sketches.
LumiMood~\cite{Oh2024-gl} is an AI that automatically adjusts lighting and post-processing; when a designer enters an initial scene and desired emotional impact, it suggests a template for a specific mood.
LLMR~\cite{Friston2021-td} is a framework for building complex virtual worlds using natural language; it interprets user input, generates code, and executes it within the Unity game engine to realize the user's vision.

DreamCodeVR~\cite{Giunchi2024-gt} achieves interactive scripting in the VR space by converting voice input into natural language and generating Unity C\# scripts via LLM. This is a similar concept to the MagicItem proposed in this study. However, the method of directly generating Unity C\# scripts has a large search space, and more sophisticated instructions are required to minimize errors while maintaining efficiency. In addition, DreamCodeVR is implemented on the Ubiq platform~\cite{Friston2021-td}, which was developed primarily for research purposes and has not been integrated with commercial metaverse platforms or subjected to large-scale user testing.

\subsection{Large Language Models for Code Generation}
Large Language Models (LLMs) are also trained from source code, and there is a great deal of interest in the source code generated by OpenAI's GPT series~\cite{Chen2021}.
There are also many LLM models developed with the goal of generating source code. One of the most successful of these is GitHub Copilot~\cite{github-copilot}, which is integrated into integrated development environments (IDEs) such as VSCode, and many developers use the automatically generated code for code completion on a daily basis.

Following the success of GitHub Copilot, many other LLMs have been built, including Amazon Q Developer~\cite{amazon_q_developer} and Tabnine~\cite{tabnine}.
These LLMs have greatly contributed to improving developer productivity, making coding more efficient, and improving quality.
In addition, LLM for code generation is becoming capable of not only complementing but also directly solving problems, such as solving competitive programming problems~\cite{Yan2023-sz}.

LLM-based code generation is also expected to be applied in the field of programming education.
For beginners, understanding the syntax and rules of programming can be a significant barrier, but the use of LLM-based code generation has the potential to make learning more efficient~\cite{Gao2015}.
In addition, a system has also been proposed that automatically generates programming problems using LLM and provides them to learners~\cite{Gulwani2017}.
Our system was inspired by these efforts and uses a mechanism that generates code from natural language alone.
Unlike code completion, which requires a certain level of programming ability to demonstrate its value, this mechanism is expected to work effectively for those without programming experience or novices.

On the other hand, there are also some problems that have been pointed out with code generation using LLM. It is difficult to guarantee the accuracy and security of the generated code, and in some cases, there is a possibility of generating code with bugs or vulnerabilities~\cite{Pearce2022}.
In addition, since the source code learned by LLM involves issues such as copyright and intellectual property rights, care must be taken when using the generated code~\cite{Allamanis2019}.
Even in interactive systems that use LLM-generated code, such as this research, these issues will need to be addressed in the future.

\section{MagicItem}
We designed MagicItem, a system that can generate the behavior of objects in VR spaces using natural language, on a commercial metaverse platform.
In this section, we discuss the design of the system, prompt engineering using LLM, and the design of large-scale online experiments.

\subsection{System Design}
We chose Cluster, a commercial metaverse platform, as the implementation platform.
Cluster is a multi-platform system that runs on standalone applications for Windows and macOS, VR headset environments, including Meta Horizon OS, and smartphone environments such as iOS and Android. 

The main reason for choosing Cluster is its implementation of Cluster Script, a JavaScript-based programming language that allows describing the behavior of objects in VR space. Also, Cluster has a feature called Script Editor. When a user specifies a particular object within the space, this editor is launched within the Cluster interface. By writing Cluster Script code in the editor, users can dynamically modify the behavior of objects in Cluster. If the generated code is incorrect, it will be displayed as an error in the console on the screen. 

In addition, the Script Editor allows the edited script to be written to a temporary file and synchronize it.
With this functionality, we implement a tool that uploads the generated code to Cluster by overwriting the temporary file with the Cluster Script generated by LLM. This tool is implemented independently from the main Cluster application and includes a distributable web server and HTML UI. Natural language input from the UI is converted into an input prompt for the LLM within the tool's web server. The prompt is sent to the OpenAI GPT-4 API and the GPT-4 API responds with a cluster script. The web server within the tool then writes the returned cluster script to the temporary file synchronized with the script editor.

Note that our system, implemented via a Cluster Script, allows code to be generated that is specialized for functions within the VR space, and object communication, synchronization, error handling, and security for code injection, can be delegated to Cluster functionality. In previous work, code had to be generated for the entire Unity C\# domain, requiring prompt engineering that considered multiple conditions related to objects and build processes, as well as code parsing and compatibility checks with the Unity environment. In contrast, the proposed tool allows for a simplified implementation by omitting the compilation pipeline.



\subsection{Prompt Engineering}
In our system, we use GPT-4 to generate code in Cluster Script format. However, due to the scarcity of online code examples for Cluster Script, it is unlikely that GPT-4 has been trained on Cluster Script. 
Also, as Cluster Script specifications are currently being extended, we need to create prompts that can accommodate these specification extensions.

To solve this problem, we used the Cluster Script definition files. Cluster Script supports TypeScript transpiling and provides the necessary TypeScript definition files. These TypeScript definition files contain sample code in the comments that demonstrate the use of the defined objects and methods. By including these definition files in the prompt, we can facilitate one-shot learning for the LLM.

By embedding the TypeScript definition file, we constructed the following prompt:

\vspace{1mm}
\begin{myverbbox}
You are a talented programmer. Please write a new function using the following JavaScript interface definition.
\vskip\baselineskip
\# Interface definition

\{CLUSTER\_SCRIPT\_DEFINITION\}

\vskip\baselineskip
\# Instructions

Please write definitions for methods that are not in the interface definition. 

Please only output the source code enclosed in
\verb|```javascript and ```|. Do not output anything other than code.
\end{myverbbox}
\vspace{1mm}

We insert the TypeScript definition file into the \verb|CLUSTER_SCRIPT_| \verb|DEFINITION| section without any modifications. The parts other than the TypeScript definition file are very concise. As a result, even if Cluster Script is extended in the future, it can be accommodated by simply modifying the inserted TypeScript definition.

\subsection{Selection of LLMs for Code Generation}
In this study, we explored the use of OpenAI's LLM models, specifically gpt-4o-2024-05-13, gpt-4-turbo-2024-04-09, and gpt-3.5-turbo-0125. While gpt-3.5-turbo offers fast response times, its input prompt is limited to 16,385 tokens, making it infeasible to include all the necessary TypeScript definitions for Cluster Script. On the other hand, gpt-4-turbo and gpt-4o both support a prompt length of 128,000 tokens, allowing the inclusion of all TypeScript definitions for Cluster Script. However, the response speed of gpt-4-turbo is slower than gpt-3.5-turbo, and the code generated by gpt-4o often did not work as intended.
Based on the above comparison, we selected gpt-4-turbo as the best balance between prompt length and code functionality.

\section{User Study}
To measure the effectiveness of our tool, we conducted a user study. The experiment was conducted entirely online. To the best of our knowledge, this is the first attempt to build an ecosystem for online VR experiments on a large-scale consumer metaverse platform. This section describes the online experiment procedure.

Note that while our system can generate code in any language supported by GPT-4, Cluster is a platform predominantly used by Japanese speakers. Therefore, the input in the natural language in this experiment was provided in Japanese.
This study was approved by Cluster, Inc. Research Ethics Committee (No. 2024-002).

\subsection{Participant Demographics}
We recruited participants online, and a total of 63 users (47 males, 10 females, six prefer not to answer) participated in the experiment.
The questionnaire also collected the frequency of use of the Cluster Script by the participants, with 26 participants having never used it, 18 participants using it approximately once a month, 10 participants using it approximately once a week, five participants using it three or more times a week and four participants using it everyday.
These participants were gathered through user-community events on Cluster and social media.

\subsection{Pre-experiment Setup}
Participants accessed the website for this experiment. The website contained an explanatory document that provided an overview of the experiment, the research purpose, methods, privacy and data handling policies, rewards, and contact information for the person in charge. Users could proceed to the subsequent experimental steps by reading all the information and checking a box to indicate their consent to participate.

To participate in the experiment, we needed to distribute the experimental items to each user's Cluster account. By submitting their Cluster API access token through the website, users had the experimental items added to their account. In this study, we distributed two items: a large white box that can be sat on (SimpleChair) and a small blue box that can be held in the hand (SimpleGrabbable). The users then downloaded the software used for the experiment. Running this software brought up an HTML page containing a form for entering natural language and a window displaying the results of the execution. While running this software, users then created a new world (a space where users can freely design within) on Cluster and placed the two items in the space. By pointing to the item for which they wanted to edit the script, users could launch the script editor within Cluster. With this Script Editor open, users typed natural language into the form in our application. The software then generated the Cluster script via the LLM and automatically rewrote the temporary file in the path where Cluster was installed, allowing users to review the generated script in the Script Editor. Finally, by pressing the Run button on Cluster, the script was applied to the item.
All user studies were conducted using the participants' personal Windows PCs or Macs.

\subsection{Tasks and Metrics}
\subsubsection{Tasks}
\begin{figure}[t]
    \centering
    \includegraphics[width=1\linewidth]{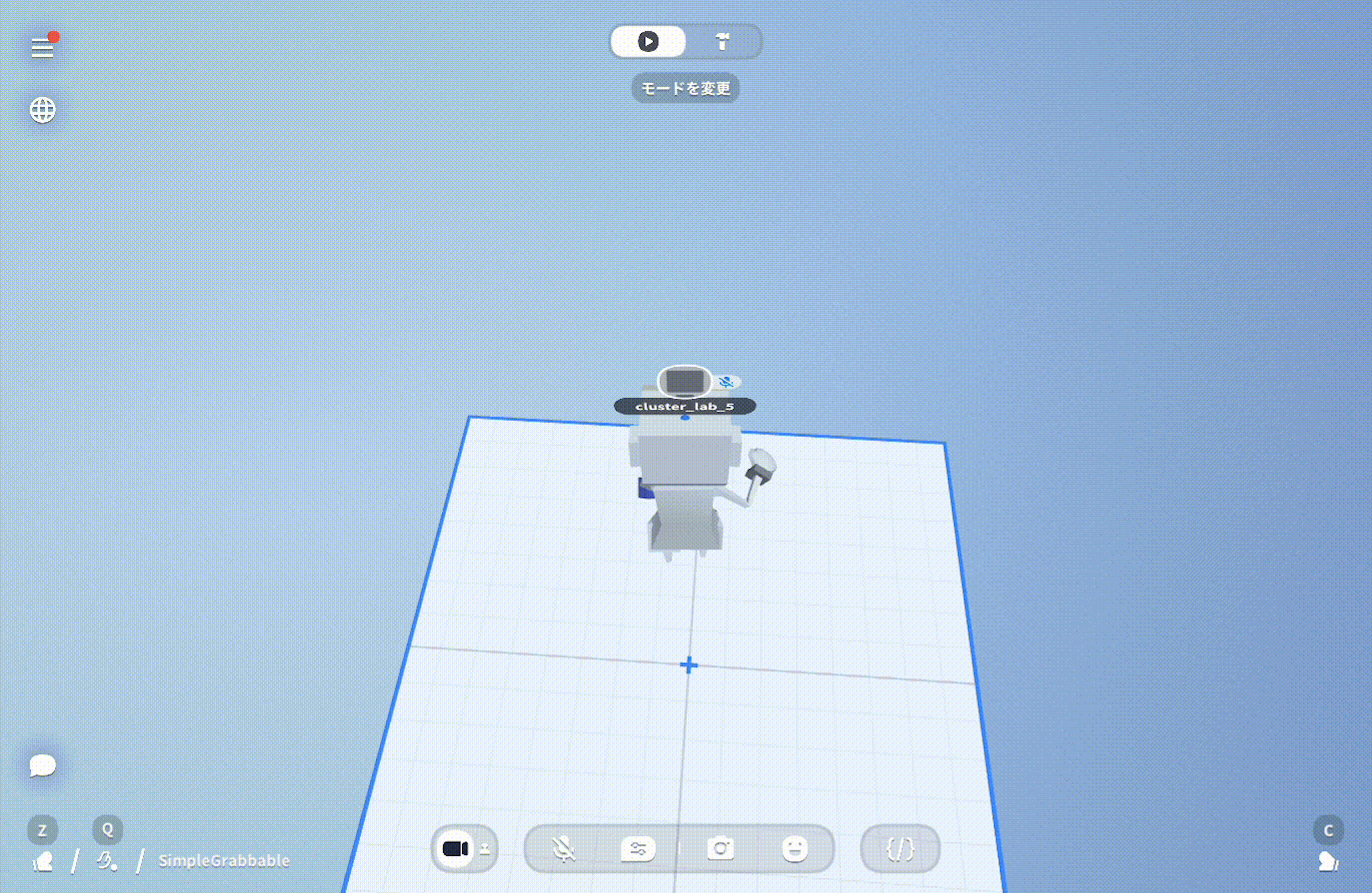}
    \caption{The goal image of Task 1: an avatar jumping higher than usual.}
    \label{fig:jumping-higer}
    \vspace{-3mm}
\end{figure}

We asked participants to perform three types of tasks. The actual tasks were not instructed using natural language, but rather by saying that users interpreted the expected behavior from the reference image/video and entered it in natural language.
The specified tasks were as follows:
\paragraph{Task 1:} We showed users a video of an avatar jumping much higher than usual (Fig.~\ref{fig:jumping-higer}), then asked them to create an object that would allow the same behavior.

\begin{figure}[t]
    \centering
    \vspace{3mm}
    \includegraphics[width=1\linewidth]{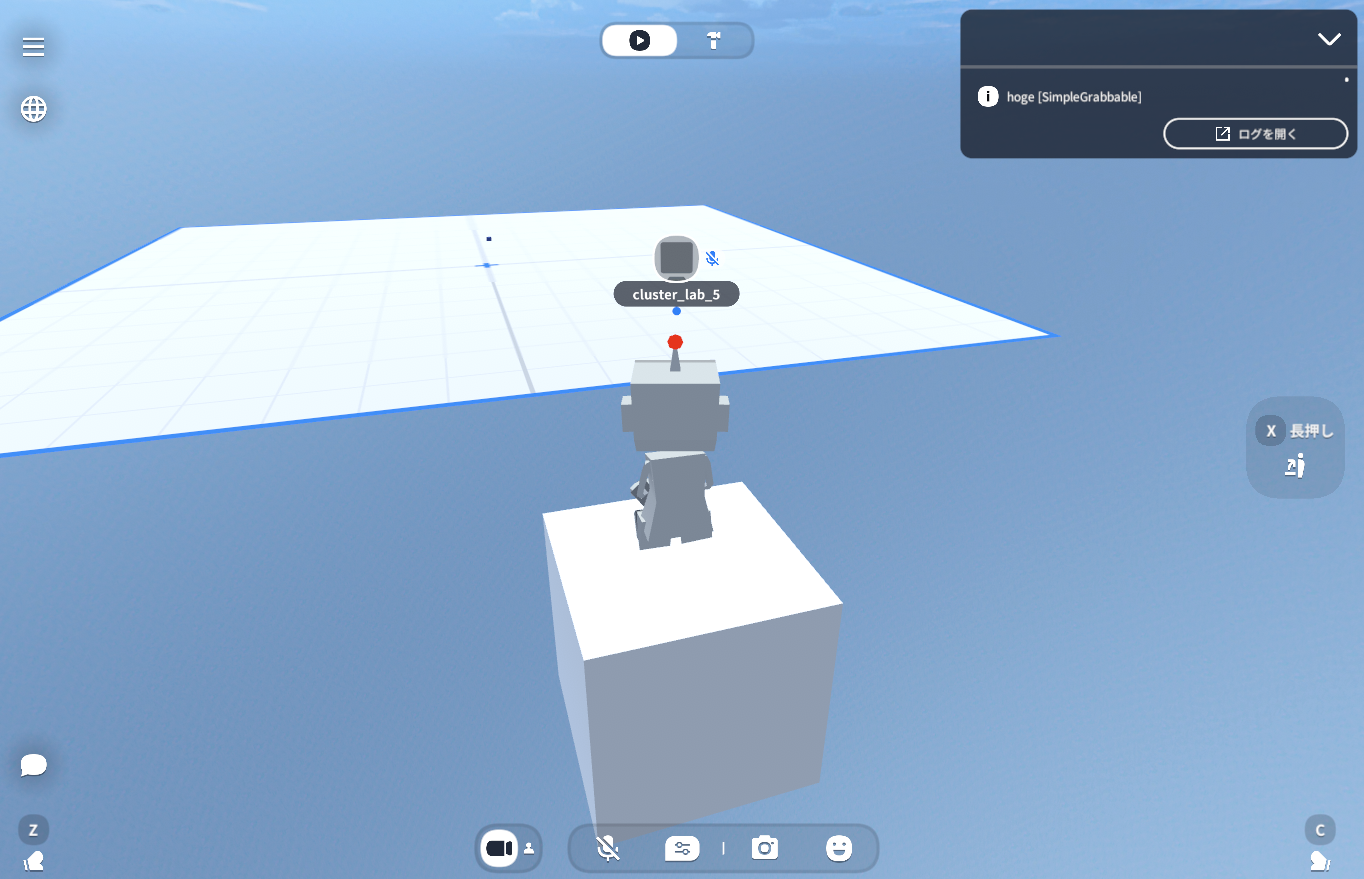}
    \caption{The goal image of Task 2: a white box floating in a position without ground, which is normally unreachable, and an avatar sitting on top of it.}
    \label{fig:without-ground}
    \vspace{-3mm}
\end{figure}

\paragraph{Task 2:} We showed users an image of a white box floating in a position without ground, which is normally unreachable, and an avatar sitting on top of it (Fig.~\ref{fig:without-ground}). We then asked them to sit on this white box and move to a space without floors.

\paragraph{Task 3:} We asked users to freely imagine and rewrite the behavior of objects as often as they wished.\\

We asked participants to press a button when starting a task on the website. This allowed us how long it took them to complete each task on Cluster. 

\subsubsection{Subjective Questionnaires}
After completing all tasks, users responded to a post-experiment questionnaire through Google Forms. The post-experiment questionnaire included the following questions:
\begin{itemize}[leftmargin=*]
    \item Whether the items were successfully created (5-point scale, 1: could not make it at all - 5: made it very well) and comments (free description) in Task 1 and Task 2, respectively.
    \item The behavior they tried to create in Task 3 (free description), whether the item in Task 3 behaved as imagined (multiple choice: behaved as intended / different than imagined, but enjoyed the creative and different behavior / different than imagined and was not enjoyable / did not work at all), and comment (free description).
    \item Usability evaluation of the entire experiment with SUS~\cite{Brooke1996-kw} questionnaire.
    \item Perception of the workload throughout the experiment with NASA-TLX~\cite{Nasa1986-vw} questionnaire. The original NASA-TLX uses a 100-point scale, but due to the limitation in Google Forms, a 10-point scale (1-10) was used, and the results of $n$ points were converted to a 100-point scale using formula $(n-1) \times 100 / 9$.
    \item Overall impressions of using the system in the experiment (free description).
\end{itemize}
Users who completed all tasks in the user study and responded to the post-experiment questionnaire were awarded 5000 Cluster Points. Cluster Points are a virtual currency that can only be used within Cluster, and 5000 points can be used to purchase user-generated items: one avatar or five avatar accessories.

\subsubsection{Quantitative Metrics}
We also internally measured user task performance using quantitative metrics:
\begin{itemize}[leftmargin=*]
\item Task completion time calculated from the interval between button presses at the end of each task.
\item Objective assessment of task success or failure extracted from logs stored on the Cluster server. In Task 1, we checked whether the y-coordinate of the user's position in VR space was greater than 2, which implies that the user has reached a height that cannot be reached normally. In Task 2, we checked whether the user's position reached the ungrounded region.
\item Number of queries per user executed by the user during each task execution.
\item Script generation time. The start time is when OpenAI API receives the request \verb|response.created|, and the end time is when OpenAI API receives the response \verb|response.completed|, which represents the time when the response is received from OpenAI and processing is complete. This generation time plus about 1 second is the total generation time, including network round-trip delays~\cite{Kurai2024-wf}.
\item Length of user prompts and length of code generated per query.
\end{itemize}

\section{Results And Analysis}

\subsection{Task Completion Times and Number of Attempts}
\begin{figure}
    \centering
    \vspace{-5mm}
    \includegraphics[width=1.0\linewidth]{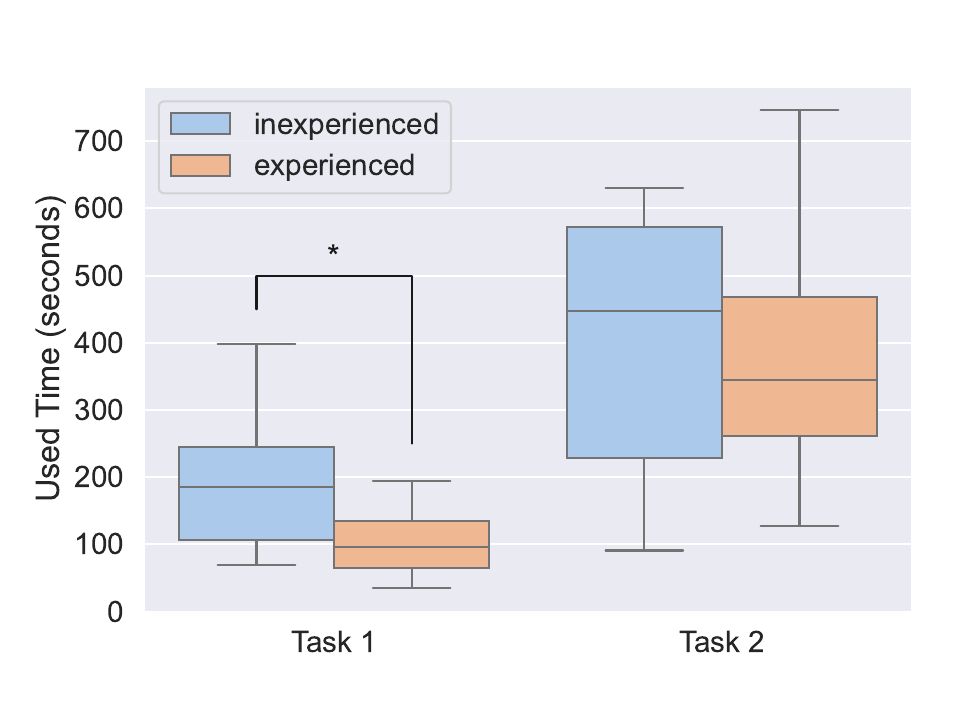}
    \caption{Completion time for each task for participants who are inexperienced and experienced Cluster Script.}
    \label{fig:completion_time}
\vspace{-3mm}
\end{figure}

Figure \ref{fig:completion_time} shows the task completion times for users. As the experiment was conducted online, some participants were suspected of leaving tasks midway, resulting in large outliers in the completion times. Therefore, we used the median as the basis for evaluation.

For participants who have no experience of Cluster Script, the median completion times were 186 seconds for Task 1 and 448 seconds for Task 2, respectively. In contrast, participants who have experience in Cluster Script complete their task in 96 seconds and 345 seconds, respectively.

Because homogeneity of variance assumption was not violated (Levene's test, \textit{p} $>$ .05),
we conducted an Mann-Whitney \textit{U} test. 
The result showed that task completion time was significantly higher for participants who have no experience of Cluster Script than for participants who have experience of Cluster Script (\textit{p} =  0.015, Cohen's \textit{r} = 0.34).

Task 3 is an exploratory task, and unlike Tasks 1 and 2, the completion time indicates how long users stayed in the room while repeatedly trying the tool. Participants who have experience of Cluster Script complete their task in 909 seconds and 1640 seconds, respectively.

\subsection{Success Rate for Task 1 and 2}
Table \ref{tab:success_ratio} shows the number of participants who succeeded in each task. 
We measured the success rate of the tasks in two ways: an objective indicator based on server logs and self-reported by the participants through the questionnaire. For self-reports, a score of four or higher was considered ``task success.''

The discrepancy between the number of server observations and self-reports is because some participants had high ideals of task performance in their self-reports. For example, although some participants objectively succeeded in the tasks, they gave themselves a score of one because the object behavior was not ideal.
Also, the number of participants in each task varies because some participants stopped during the task, and we could not get their exact server logs.


\begin{table}[b]
\centering
\caption{The success rate for each task based on server log and self-report. For Task 1 and Task 2, the count participants for whom the server log was accurately recorded. For Task 3, the number of self-reports is counted as the number of participants.}
\begin{tabular}{cccc}
\hline
& Task1          & Task2          & Task3        \\ \hline
\# of participants                                                    & 58             & 45             & 63           \\ \hline
\begin{tabular}[c]{@{}c@{}}Success rate\\ (server log)\end{tabular}   & 86.2\% (50/58) & 100\% (45/45)  & -            \\ \hline
\begin{tabular}[c]{@{}c@{}}Success rate\\ (self-reports)\end{tabular} & 84.5\% (49/58) & 73.3\% (33/45) & 23.7\% (22/63) \\ \hline
\end{tabular}
\label{tab:success_ratio}
\vspace{-3mm}
\end{table}

\subsection{Attempted Behaviors in Task 3}
The behaviors attempted in the free-form Task 3 can be summarized as follows:
\begin{enumerate}[leftmargin=*]
    \item \textit{Specify movements:} repetitive movement, circular motion, rotation, up and down movement, random movements, etc.
    \item \textit{Enhance user actions:} reduced falling speed, ascent, increased jumping power, changed movement speed, etc.
    \item \textit{Vehicle-type items:} controllable vehicles, roller coaster-like movements, elevator-like ascent, etc.
    \item \textit{Biological movements:} following the player like a pet, moving freely like a bird, etc.
    \item \textit{Change physical laws:} moon gravity while being held, ignoring gravity, or floating in the air, etc.
    \item \textit{React specific actions:} reacting when waved, flying away when released, bouncing when hitting the ground, etc.
    \item \textit{Change the environment:} darkening the room while being held, applying post-processing color filters, etc.
\end{enumerate}

From the questionnaire, (1-3) often performed as expected. (4-5) sometimes differed from expectations, but users enjoyed the behaviors. (6) had unexpected movements that users did not enjoy. (7) is not currently supported by Cluster Script, and errors occurred in all cases. In addition, attempts to explore the capabilities of LLMs were observed, such as telling them to ``do something interesting.''

\subsection{Usability and Perceived Workload Evaluation}
\paragraph{Usability.} The overall mean SUS score was 62.
When categorized by the frequency with which participants used Cluster Script, the group that had never used it had an average SUS score of 66, while the group that used it once a month or more had an average SUS score of 59.
We tested for significant differences between these two groups.
Since normality (Shapiro-Wilk test, \textit{p} $>$ 0.05) and equal variance (Levene test, \textit{p} $>$ 0.05) were not rejected, we performed a Student's \textit{t} test.
The results showed that the SUS was marginally significant higher for the group that never used Cluster Script compared to the group that used it once a month or more (\textit{t} = 1.84, \textit{p} = 0.07).

\begin{figure}
    \centering
    \includegraphics[width=1\linewidth]{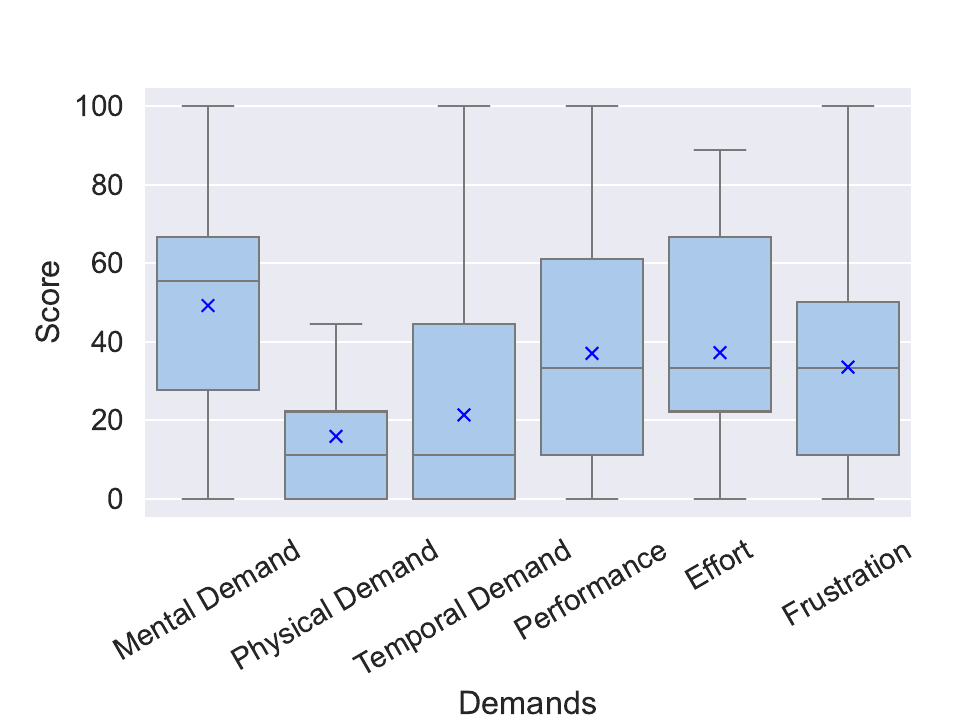}
    \caption{Scores for each of the six fundamental items used to assess task workload in the NASA-TLX.}
    \label{fig:nasa-tlx}
\end{figure}

\paragraph{Perceived workload.}
Figure \ref{fig:nasa-tlx} shows the mean and standard deviations for each item in the NASA-TLX questionnaire collected from the participants. Mental Demand: 49.21~$\pm$24.50,
Physical Demand: 15.87~$\pm$20.81,
Temporal Demand: 21.34~$\pm$24.22,
Performance: 37.04~$\pm$28.43,
Effort: 37.21~$\pm$27.23 and
Frustration: 33.51~$\pm$25.67.

\subsection{Metrics of Code Generation}
\subsubsection{The Number of Attempts}
Figure~\ref{fig:code-generation-analysis} (A) shows the number of trials for each participant.
When the groups were divided into those who were completely inexperienced and those who were experienced in scripting, the median number of trials for the inexperienced participants was 1, 3, and 9 for Tasks 1, 2, and 3, respectively, while the median number of trials for the experienced participants was 1, 3, and 3, respectively.
Because the homogeneity of variance assumption was not violated (Levene's test, \textit{p} $>$ .05),
we conducted an Mann-Whitney \textit{U} test. 
For Task 3, the result showed that the number of attempts was significantly lower for participants who have no experience of Cluster Script than for participants who have experience of Cluster Script (\textit{p} = 0.028, Cohen's \textit{r} = 0.29).

\subsubsection{Code Generation Time}
Figure~\ref{fig:code-generation-analysis} (B) shows the time taken to generate scripts.
The median time of trials for the inexperienced participants was 5, 7, and 10 seconds for Tasks 1, 2, and 3, respectively, while the median number of trials for the experienced participants was 6, 8, and 12, respectively.

Because the homogeneity of variance assumption was not violated (Levene's test, \textit{p} $>$ .05),
we conducted an Mann-Whitney \textit{U} test. 
For Task 3, the result showed that the code generation time was significantly lower for participants who have no experience of Cluster Script than for participants who have experience of Cluster Script (\textit{p} = 0.0087, Cohen's \textit{r} = 0.13).


\subsubsection{Length of User Prompt}
Figure~\ref{fig:code-generation-analysis} (C) shows the length of the instructions entered by the participants to generate the Cluster Scripts. The length is measured by the amount of tokens in GPT-4-turbo. 
The median token length of trials for the inexperienced participants was 104, 111, and 218 for Tasks 1, 2, and 3, respectively, while the median token length for the experienced participants was 99, 103, and 340, respectively.

Since the assumption of homogeneity of variance was not violated (Levene's test, \textit{p} $>$ .05), we conducted a Mann-Whitney \textit{U} test. For Task 3, the result showed that the length of the user prompt was significantly shorter for participants without Cluster Script experience than for participants with Cluster Script experience (\textit{p} = 0.037, Cohen's \textit{r} = 0.10).




\subsubsection{Length of Generated Code}
\begin{figure*}
    \centering
    \includegraphics[width=1\linewidth]{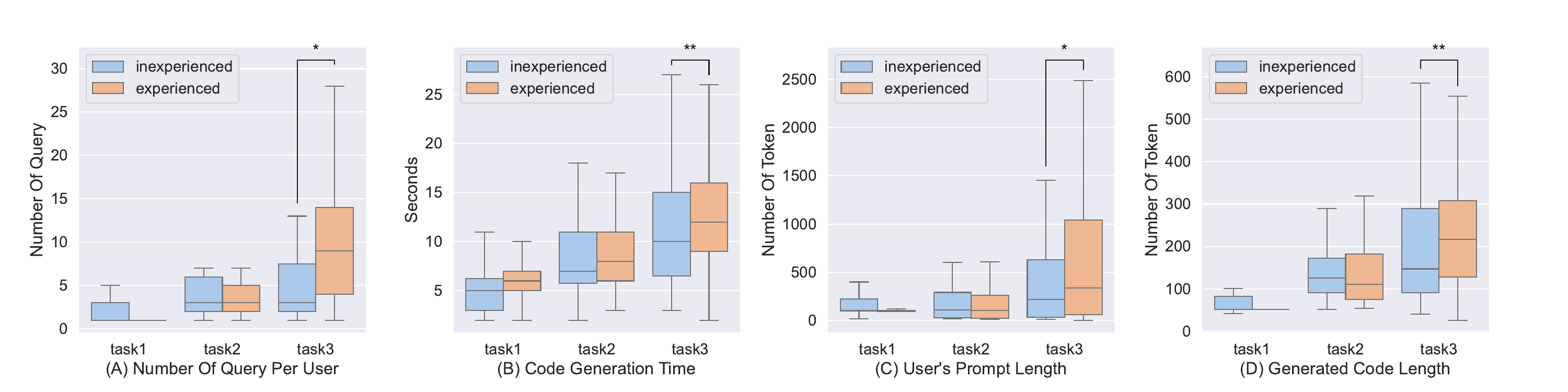}
    \caption{The figure (A) shows the number of trials for each participant, (B) shows the time taken to generate scripts, (C) shows the length of the instructions entered by the participants to generate the Cluster Scripts, and (D) measures the length by the quantity of tokens in GPT-4-turbo.}
    \label{fig:code-generation-analysis}
\end{figure*}

Figure~\ref{fig:code-generation-analysis} (D) shows the length of the Cluster Script generated by LLM. The length is measured by the amount of tokens in GPT-4-turbo.
The median token length of trials for the inexperienced participants was 51, 126, and 147 for Tasks 1, 2, and 3, respectively, while the median token length for the experienced participants was 51, 110, and 217, respectively.



Because the homogeneity of variance assumption was not violated (Levene's test, \textit{p} $>$ .05),
we conducted an Mann-Whitney \textit{U} test. 
For Task 3, the result showed that the length of generated code was significantly lower for participants who have no experience of Cluster Script than for participants who have experience of Cluster Script (\textit{p} = 0.0082, Cohen's \textit{r} = 0.13).

From the above values, we can see that a longer Cluster Script is generated for Task 2 than for Task 1, and for Task 3 than for Task 2. We can also see that the Script generation time is also longer.

\subsection{Free-form Description for Each Task}
For free descriptions of each task, the following responses were obtained:

\paragraph{Task 1:} Responses were collected from 39 participants, with many expressing positive opinions such as ``easy to do,'' ``smooth to do,'' ``worked accurately,'' and ``raised expectations for the system.'' Some participants reported that although errors occurred or the desired behavior was not achieved on the first try, they could achieve the expected results by trying again or breaking down the phenomena into smaller descriptions. On the other hand, there were a few cases where participants found it difficult to handle detailed processes such as gravity or flight distance, resulting in situations such as flying endlessly and not being able to land properly.

\paragraph{Task 2:} Responses were collected from 40 participants, with many reporting that they could complete the task without any particular problems. However, experiences with unexpected behavior, such as errors or moving to extremely distant locations, were also reported. Some participants felt that detailed prompts were necessary, as instructions such as ``move to a region with no floor'' or ``move forward'' did not work well. A few participants attempted to implement key input-based motion by treating the object as a vehicle but were unsuccessful and settled for simple motion.

\paragraph{Task 3:} Responses were collected from 46 participants. Many participants, especially those without programming experience, expressed the ability and enjoyment of being able to command movements with words without the need for scripting knowledge, such as ``It feels like the area I can control has expanded, which is fun'' and ``Imagining being able to create gimmicks in the future just by giving commands is exciting!'' On the other hand, many participants pointed out the skills and knowledge required of users, such as understanding 3D world-specific keywords, Cluster Scripts specifications, writing efficient scripts, language expression skills, and customizing generated results. Participants familiar with scripting noted that the difficulty of implementing complex motion and the ambiguity of natural language input.

\section{Discussion}
\paragraph{Task completion time and number of attempts.} Existing studies have assigned tasks such as moving an item to a specified location or having items perform certain actions with each other. In contrast, this study aims to generate code that involves user interaction, such as performing certain operations while the user holds or sits on an item. Nevertheless, in Task 1, most users completed the task on the first attempt, and in Task 2, more than half of the users completed the task after about four trial-and-error attempts. This demonstrates the effectiveness of generating code in the cluster script domain, which is closely tied to the platform, and shows that our system can intuitively generate more advanced behaviors that involve interaction.
We assume that there is not much difference in the median task completion time for Task 1 and Task 2 between inexperienced and experienced users.
This suggests that both inexperienced and experienced users can perform creative tasks quickly using our system.

\paragraph{Expressiveness.}
The expressiveness of our system is particularly evident in the various user-driven creative attempts seen in Task 3. Interestingly, it is possible to define behaviors using metaphors of real-world object motion without specifying physical motion, and some of these can be implemented as the natural language input described. For example, the prompt ``make the item's gravity like that of the moon only while holding the item'' requires the LLM to have general knowledge and interpretation of ``moon's gravity'' to generate the script, but our tool successfully interpreted the prompt and generated this script.
Also, by giving the instruction ``You are a bird,'' it was possible to make the object fly like a bird. This is an example of how complex movements can be achieved with short, descriptive sentences.

Participants commented that it was convenient to be able to write the code without having to remember the API names or the order of the arguments, and that it was interesting to be able to incorporate actions into items without having to know anything about scripting. There were also comments that it felt like a sudden growth in the sense that the range of things you can control has expanded, and find it fun and surprising that they could manipulate objects using natural language input.
In contrast, there were also comments that some knowledge of Unity is necessary, even if you don't have knowledge of scripting, and that it is difficult to try to combine multiple actions. When trying to make complex actions, it may be necessary to have knowledge of Unity, the action environment, or to be creative in the way you give instructions. 

\paragraph{Usability.}
SUS scores are higher for those without Cluster Script experience, which suggests that the system is easy to use for novices but difficult for experienced users.
We believe that this perception is due to differences in the level of expectation and the points of expectation between novice and experienced users.
Analyzing the results of the free-response questionnaire, novices express surprise and positive evaluations of the system due to the change in 3D object behavior without programming knowledge.
In contrast, experienced users seem to feel frustrated when they realize what they could do if they wrote the code themselves and experience the inability to immediately write accurate behaviors due to the LLM intermediary.

Note that in the current evaluation setup, the overall impression of the experiment, including the installation of additional tools and the operation of the cluster application, became the target of the SUS evaluation.
Our method requires multiple steps to install the tool and requires the HTML page where the prompt is entered and the Cluster application to be opened simultaneously, and this implementation may have lowered the SUS score.

\paragraph{Perceived Workload.}
The NASA-TLX score results showed that the median scores for Physical Demand and Temporal Demand were relatively low (below 25). In particular, the trend for low Physical Demand scores was similar to that found in the previous study in DreamCodeVR~\cite{Giunchi2024-gt}.
On the other hand, for Temporal Demand, while the average DreamCodeVR scores are between 37 and 75\footnote{In the DreamCodeVR experiment, the NASA-TLX score was listed on a 20 point scale, so to compare it to our 100 point scale, the scores were multiplied by 5.}, our MagicItem system seems to show a lower value, although it is not possible to perform a statistical analysis.
This difference may be due to the fact that MagicItem uses direct text input, while DreamCodeVR uses voice input. Speech input takes a certain amount of time to process, so it may be that users felt a greater temporal demand while waiting for the system to respond.

Note that the results of this experiment showed a very high degree of general variability.
One possible factor is that the NASA-TLX questions are worded and difficult to understand for general users outside of the research community, leading to confusion.
In fact, some of the participants responded that they did not understand the meaning of the questions.


\section{Limitations and Future Work}
We asked for general impressions and feedback at the end of the questionnaire and received 55 responses.
The responses reveal the remaining limitations and directions for future research on MagicItem:

\paragraph{Usability and Accessibility.} 
While the system is intuitive for some users, especially those without programming experience, others find it challenging to use effectively without prior knowledge of scripting or programming concepts.
In contrast, advanced users also found it difficult to perform accurate scripting. User comments have suggested providing more guidance, such as sample prompts and explanations tailored to different skill levels (beginner, intermediate, advanced), to make the system more accessible to a wider range of users.
As development experience within Cluster is gained from the server, it may be possible to fine-tune personalized code generation LLMs based on user skill level and preferences.

\paragraph{Error Handling and Debugging.} 
Users often encountered errors and unexpected behavior when generating scripts that required knowledge of the Cluster Script to diagnose and resolve.
Currently, the Cluster Script definition itself is used as an input prompt, and features such as automatic compilation error repair are not implemented.
Improving the system's error-handling capabilities, providing more informative error messages, and offering debugging suggestions could greatly improve the user experience and reduce frustration.

\paragraph{Prompt Engineering and Natural Language Understanding.} 
The effectiveness of generated scripts depends heavily on the user's ability to provide clear and well-structured prompts.
Although it is ideal for the LLM itself to have better natural language understanding capabilities, providing guidance on effective prompt engineering techniques could help users achieve their desired results more consistently.

\paragraph{Integration and Workflow.}
The current workflow requires participants to navigate between three environments: the Cluster application, the code generation tool using LLMs, and the web server explaining the experimental procedure. Some users found this complexity cumbersome and confusing. This complexity also made it difficult to integrate participant behavior across the three environments from a measurement perspective. For example, when participants performed each task, there was no way to combine information such as the coordinates of the participant in Cluster and the prompt sent to the LLM except by comparing timestamps. As a result, the correct information could not be combined unless participants followed the experimental procedure and read all of the instructions before performing the tasks. Streamlining this workflow and integrating all systems more seamlessly into the Cluster platform is expected to improve the user experience, as indicated by the SUS scores.

\paragraph{Authoring the Entire VR Space.}
In Task 3, users expected to be able to edit aspects beyond behavior, such as shaders and materials, using LLMs. In the area of LLMs, there is active research on various multimodal outputs from natural language, including 3D modeling, texture, animation, and audio generation. Methods for integrating these multimodal generation models to create VR spaces have also been proposed, but are currently limited to offline editing. Research on interactive editing within VR spaces is expected.

\paragraph{Collaboration and Sharing.} 
Users have expressed interest in collaborating with others and sharing their creations within the Cluster community. In this experiment, participants had to prepare their own Cluster spaces to distribute craft items, making it difficult to gather multiple users in the same space. To conduct experiments that evaluate collaborative tasks, it is necessary to develop platform-side arrangements, such as delegating item editing rights among participants. In addition, user feedback suggests that features that facilitate collaboration, such as shared workspaces, version control, and easy sharing of scripts and behaviors, could foster a vibrant ecosystem of user-generated content and knowledge sharing.

\paragraph{Advanced Functionality and Flexibility.}
The current Script Editor excels at creating behaviors within a single object, but users have expressed interest in creating more complex interactions, such as communication between multiple objects and collaborative behaviors. Extending the system's capabilities to support more advanced scripting scenarios could open up new possibilities for user creativity and engagement.

\section{Conclusion}
This study introduces MagicItem, a novel system that integrates LLMs with Cluster, allowing users, even those without programming experience, to generate interactive object behaviors using natural language. 
We conducted the first large-scale online experiment on the consumer metaverse platform with 63 participants with varying levels of scripting expertise. The results demonstrated the effectiveness and usability of MagicItem, with the majority of participants completing the object behavior modification tasks.
Usability tests showed that novice users rated the system higher than experienced users. A free-form task confirmed the rich expressiveness of the system, with participants implementing a variety of object behaviors.
The qualitative analysis of user feedback provided insight into key challenges and future research directions.

Our MagicItem represents a significant step toward democratizing content creation in metaverse platforms, potentially opening up new forms of creativity and engagement for users from diverse backgrounds. As LLMs rapidly evolve, the potential applications of systems such as MagicItem are vast. By combining natural language processing with the immersive and social nature of metaverse platforms, we can create more accessible, engaging, and collaborative virtual experiences. This research serves as a foundation for future developments, paving the way for a new era of user-generated content and interactive experiences in the metaverse.

\begin{acks}
This work was partially supported by JST ASPIRE Grant Number JPMJAP2327.
\end{acks}

\bibliographystyle{ACM-Reference-Format}
\bibliography{magicitem}


\begin{thebibliography}{41}


\ifx \showCODEN    \undefined \def \showCODEN     #1{\unskip}     \fi
\ifx \showDOI      \undefined \def \showDOI       #1{#1}\fi
\ifx \showISBNx    \undefined \def \showISBNx     #1{\unskip}     \fi
\ifx \showISBNxiii \undefined \def \showISBNxiii  #1{\unskip}     \fi
\ifx \showISSN     \undefined \def \showISSN      #1{\unskip}     \fi
\ifx \showLCCN     \undefined \def \showLCCN      #1{\unskip}     \fi
\ifx \shownote     \undefined \def \shownote      #1{#1}          \fi
\ifx \showarticletitle \undefined \def \showarticletitle #1{#1}   \fi
\ifx \showURL      \undefined \def \showURL       {\relax}        \fi
\providecommand\bibfield[2]{#2}
\providecommand\bibinfo[2]{#2}
\providecommand\natexlab[1]{#1}
\providecommand\showeprint[2][]{arXiv:#2}

\bibitem[Allamanis(2019)]%
        {Allamanis2019}
\bibfield{author}{\bibinfo{person}{Miltiadis Allamanis}.} \bibinfo{year}{2019}\natexlab{}.
\newblock \showarticletitle{The adverse effects of code duplication in machine learning models of code}. In \bibinfo{booktitle}{\emph{Proceedings of the 2019 ACM SIGPLAN International Symposium on New Ideas, New Paradigms, and Reflections on Programming and Software}} \emph{(\bibinfo{series}{Onward! 2019})}. \bibinfo{pages}{143–153}.
\newblock
\urldef\tempurl%
\url{https://doi.org/10.1145/3359591.3359735}
\showDOI{\tempurl}


\bibitem[{Amazon.com, Inc.}(2024)]%
        {amazon_q_developer}
\bibfield{author}{\bibinfo{person}{{Amazon.com, Inc.}}} \bibinfo{year}{2024}\natexlab{}.
\newblock \bibinfo{title}{Amazon {Q} Developer}.
\newblock \bibinfo{howpublished}{\url{https://aws.amazon.com/q/developer/}}.
\newblock
\newblock
\shownote{Accessed: 2024-6-15}.


\bibitem[Biocca et~al\mbox{.}(2003)]%
        {Biocca2003}
\bibfield{author}{\bibinfo{person}{Frank Biocca}, \bibinfo{person}{Chad Harms}, {and} \bibinfo{person}{Judee~K. Burgoon}.} \bibinfo{year}{2003}\natexlab{}.
\newblock \showarticletitle{Toward a more robust theory and measure of social presence: review and suggested criteria}.
\newblock \bibinfo{journal}{\emph{Presence: Teleoperators and Virtual Environments}} \bibinfo{volume}{12}, \bibinfo{number}{5} (\bibinfo{date}{oct} \bibinfo{year}{2003}), \bibinfo{pages}{456–480}.
\newblock
\showISSN{1054-7460}
\urldef\tempurl%
\url{https://doi.org/10.1162/105474603322761270}
\showDOI{\tempurl}


\bibitem[Blackwell et~al\mbox{.}(2019)]%
        {Blackwell2019}
\bibfield{author}{\bibinfo{person}{Lindsay Blackwell}, \bibinfo{person}{Nicole Ellison}, \bibinfo{person}{Natasha Elliott-Deflo}, {and} \bibinfo{person}{Raz Schwartz}.} \bibinfo{year}{2019}\natexlab{}.
\newblock \showarticletitle{Harassment in Social VR: Implications for Design}. In \bibinfo{booktitle}{\emph{Proceedings of the 2019 IEEE Conference on Virtual Reality and 3D User Interfaces}} \emph{(\bibinfo{series}{IEEE VR '19})}. \bibinfo{pages}{854--855}.
\newblock
\urldef\tempurl%
\url{https://doi.org/10.1109/VR.2019.8798165}
\showDOI{\tempurl}


\bibitem[{Blender Foundation}(2024)]%
        {blender}
\bibfield{author}{\bibinfo{person}{{Blender Foundation}}.} \bibinfo{year}{2024}\natexlab{}.
\newblock \bibinfo{title}{Blender}.
\newblock \bibinfo{howpublished}{\url{https://www.blender.org/}}.
\newblock
\newblock
\shownote{Accessed: 2024-6-15}.


\bibitem[Brooke(1996)]%
        {Brooke1996-kw}
\bibfield{author}{\bibinfo{person}{John Brooke}.} \bibinfo{year}{1996}\natexlab{}.
\newblock \showarticletitle{{SUS}: A 'quick and dirty' usability scale}.
\newblock In \bibinfo{booktitle}{\emph{Usability Evaluation In Industry}}. \bibinfo{publisher}{CRC Press}, \bibinfo{pages}{207--212}.
\newblock
\urldef\tempurl%
\url{http://dx.doi.org/10.1201/9781498710411-35}
\showURL{%
\tempurl}


\bibitem[Chen et~al\mbox{.}(2021)]%
        {Chen2021}
\bibfield{author}{\bibinfo{person}{Mark Chen}, \bibinfo{person}{Jerry Tworek}, \bibinfo{person}{Heewoo Jun}, \bibinfo{person}{Qiming Yuan}, \bibinfo{person}{Henrique~Ponde de Oliveira~Pinto}, \bibinfo{person}{Jared Kaplan}, \bibinfo{person}{Harri Edwards}, \bibinfo{person}{Yuri Burda}, \bibinfo{person}{Nicholas Joseph}, \bibinfo{person}{Greg Brockman}, \bibinfo{person}{Alex Ray}, \bibinfo{person}{Raul Puri}, \bibinfo{person}{Gretchen Krueger}, \bibinfo{person}{Michael Petrov}, \bibinfo{person}{Heidy Khlaaf}, \bibinfo{person}{Girish Sastry}, \bibinfo{person}{Pamela Mishkin}, \bibinfo{person}{Brooke Chan}, \bibinfo{person}{Scott Gray}, \bibinfo{person}{Nick Ryder}, \bibinfo{person}{Mikhail Pavlov}, \bibinfo{person}{Alethea Power}, \bibinfo{person}{Lukasz Kaiser}, \bibinfo{person}{Mohammad Bavarian}, \bibinfo{person}{Clemens Winter}, \bibinfo{person}{Philippe Tillet}, \bibinfo{person}{Felipe~Petroski Such}, \bibinfo{person}{Dave Cummings}, \bibinfo{person}{Matthias Plappert}, \bibinfo{person}{Fotios Chantzis},
  \bibinfo{person}{Elizabeth Barnes}, \bibinfo{person}{Ariel Herbert-Voss}, \bibinfo{person}{William~Hebgen Guss}, \bibinfo{person}{Alex Nichol}, \bibinfo{person}{Alex Paino}, \bibinfo{person}{Nikolas Tezak}, \bibinfo{person}{Jie Tang}, \bibinfo{person}{Igor Babuschkin}, \bibinfo{person}{Suchir Balaji}, \bibinfo{person}{Shantanu Jain}, \bibinfo{person}{William Saunders}, \bibinfo{person}{Christopher Hesse}, \bibinfo{person}{Andrew~N. Carr}, \bibinfo{person}{Jan Leike}, \bibinfo{person}{Josh Achiam}, \bibinfo{person}{Vedant Misra}, \bibinfo{person}{Evan Morikawa}, \bibinfo{person}{Alec Radford}, \bibinfo{person}{Matthew Knight}, \bibinfo{person}{Miles Brundage}, \bibinfo{person}{Mira Murati}, \bibinfo{person}{Katie Mayer}, \bibinfo{person}{Peter Welinder}, \bibinfo{person}{Bob McGrew}, \bibinfo{person}{Dario Amodei}, \bibinfo{person}{Sam McCandlish}, \bibinfo{person}{Ilya Sutskever}, {and} \bibinfo{person}{Wojciech Zaremba}.} \bibinfo{year}{2021}\natexlab{}.
\newblock \showarticletitle{Evaluating Large Language Models Trained on Code}.
\newblock  (\bibinfo{year}{2021}).
\newblock
\showeprint[arxiv]{2107.03374}~[cs.LG]


\bibitem[{Cluster, Inc.}(2024)]%
        {clsuter}
\bibfield{author}{\bibinfo{person}{{Cluster, Inc.}}} \bibinfo{year}{2024}\natexlab{}.
\newblock \bibinfo{title}{Cluster}.
\newblock \bibinfo{howpublished}{\url{https://cluster.mu/}}.
\newblock
\newblock
\shownote{Accessed: 2024-6-15}.


\bibitem[{Epic Games, Inc.}(2024a)]%
        {fortnite}
\bibfield{author}{\bibinfo{person}{{Epic Games, Inc.}}} \bibinfo{year}{2024}\natexlab{a}.
\newblock \bibinfo{title}{Fortnite}.
\newblock \bibinfo{howpublished}{\url{https://www.fortnite.com/}}.
\newblock
\newblock
\shownote{Accessed: 2024-6-15}.


\bibitem[{Epic Games, Inc.}(2024b)]%
        {unreal-engine}
\bibfield{author}{\bibinfo{person}{{Epic Games, Inc.}}} \bibinfo{year}{2024}\natexlab{b}.
\newblock \bibinfo{title}{Unreal Engine}.
\newblock \bibinfo{howpublished}{\url{https://www.unrealengine.com/}}.
\newblock
\newblock
\shownote{Accessed: 2024-6-15}.


\bibitem[Fabri et~al\mbox{.}(1999)]%
        {Fabri1999}
\bibfield{author}{\bibinfo{person}{Marc Fabri}, \bibinfo{person}{David~J. Moore}, {and} \bibinfo{person}{Dave~J. Hobbs}.} \bibinfo{year}{1999}\natexlab{}.
\newblock \showarticletitle{The Emotional Avatar: Non-verbal Communication Between Inhabitants of Collaborative Virtual Environments}. In \bibinfo{booktitle}{\emph{Gesture-Based Communication in Human-Computer Interaction}}, \bibfield{editor}{\bibinfo{person}{Annelies Braffort}, \bibinfo{person}{Rachid Gherbi}, \bibinfo{person}{Sylvie Gibet}, \bibinfo{person}{Daniel Teil}, {and} \bibinfo{person}{James Richardson}} (Eds.). \bibinfo{publisher}{Springer Berlin Heidelberg}, \bibinfo{pages}{269--273}.
\newblock
\showISBNx{978-3-540-46616-1}


\bibitem[Freeman et~al\mbox{.}(2020)]%
        {Freeman2020}
\bibfield{author}{\bibinfo{person}{Guo Freeman}, \bibinfo{person}{Samaneh Zamanifard}, \bibinfo{person}{Divine Maloney}, {and} \bibinfo{person}{Alexandra Adkins}.} \bibinfo{year}{2020}\natexlab{}.
\newblock \showarticletitle{My Body, My Avatar: How People Perceive Their Avatars in Social Virtual Reality}. In \bibinfo{booktitle}{\emph{Extended Abstracts of the 2020 CHI Conference on Human Factors in Computing Systems}} \emph{(\bibinfo{series}{CHI EA '20})}. \bibinfo{pages}{1–8}.
\newblock
\showISBNx{9781450368193}
\urldef\tempurl%
\url{https://doi.org/10.1145/3334480.3382923}
\showDOI{\tempurl}


\bibitem[Friston et~al\mbox{.}(2021)]%
        {Friston2021-td}
\bibfield{author}{\bibinfo{person}{Sebastian~J Friston}, \bibinfo{person}{Ben~J Congdon}, \bibinfo{person}{David Swapp}, \bibinfo{person}{Lisa Izzouzi}, \bibinfo{person}{Klara Brandst{\"a}tter}, \bibinfo{person}{Daniel Archer}, \bibinfo{person}{Otto Olkkonen}, \bibinfo{person}{Felix~Johannes Thiel}, {and} \bibinfo{person}{Anthony Steed}.} \bibinfo{year}{2021}\natexlab{}.
\newblock \showarticletitle{Ubiq: A System to Build Flexible Social Virtual Reality Experiences}. In \bibinfo{booktitle}{\emph{Proceedings of the 27th {ACM} Symposium on Virtual Reality Software and Technology}} \emph{(\bibinfo{series}{VRST '21})}. Article \bibinfo{articleno}{6}, \bibinfo{numpages}{11}~pages.
\newblock


\bibitem[Gao et~al\mbox{.}(2015)]%
        {Gao2015}
\bibfield{author}{\bibinfo{person}{Tong Gao}, \bibinfo{person}{Mira Dontcheva}, \bibinfo{person}{Eytan Adar}, \bibinfo{person}{Zhicheng Liu}, {and} \bibinfo{person}{Karrie~G. Karahalios}.} \bibinfo{year}{2015}\natexlab{}.
\newblock \showarticletitle{DataTone: Managing Ambiguity in Natural Language Interfaces for Data Visualization}. In \bibinfo{booktitle}{\emph{Proceedings of the 28th Annual ACM Symposium on User Interface Software \& Technology}} \emph{(\bibinfo{series}{UIST '15})}. \bibinfo{pages}{489–500}.
\newblock
\showISBNx{9781450337793}
\urldef\tempurl%
\url{https://doi.org/10.1145/2807442.2807478}
\showDOI{\tempurl}


\bibitem[{GitHub, Inc.}(2024)]%
        {github-copilot}
\bibfield{author}{\bibinfo{person}{{GitHub, Inc.}}} \bibinfo{year}{2024}\natexlab{}.
\newblock \bibinfo{title}{{GitHub} Copilot}.
\newblock \bibinfo{howpublished}{\url{https://copilot.github.com/}}.
\newblock
\newblock
\shownote{Accessed: 2024-6-15}.


\bibitem[Giunchi et~al\mbox{.}(2024)]%
        {Giunchi2024-gt}
\bibfield{author}{\bibinfo{person}{Daniele Giunchi}, \bibinfo{person}{Nels Numan}, \bibinfo{person}{Elia Gatti}, {and} \bibinfo{person}{Anthony Steed}.} \bibinfo{year}{2024}\natexlab{}.
\newblock \showarticletitle{{DreamCodeVR}: Towards Democratizing Behavior Design in Virtual Reality with {Speech-Driven} Programming}. In \bibinfo{booktitle}{\emph{2024 {IEEE} Conference Virtual Reality and {3D} User Interfaces ({VR})}}. \bibinfo{publisher}{IEEE}, \bibinfo{pages}{579--589}.
\newblock


\bibitem[Gulwani et~al\mbox{.}(2017)]%
        {Gulwani2017}
\bibfield{author}{\bibinfo{person}{Sumit Gulwani}, \bibinfo{person}{Oleksandr Polozov}, {and} \bibinfo{person}{Rishabh Singh}.} \bibinfo{year}{2017}\natexlab{}.
\newblock \showarticletitle{Program Synthesis}.
\newblock \bibinfo{journal}{\emph{Foundations and Trends® in Programming Languages}} \bibinfo{volume}{4}, \bibinfo{number}{1-2} (\bibinfo{year}{2017}), \bibinfo{pages}{1--119}.
\newblock
\showISSN{2325-1107}
\urldef\tempurl%
\url{https://doi.org/10.1561/2500000010}
\showDOI{\tempurl}


\bibitem[Hindmarsh et~al\mbox{.}(2000)]%
        {Hindmarsh2000}
\bibfield{author}{\bibinfo{person}{Jon Hindmarsh}, \bibinfo{person}{Mike Fraser}, \bibinfo{person}{Christian Heath}, \bibinfo{person}{Steve Benford}, {and} \bibinfo{person}{Chris Greenhalgh}.} \bibinfo{year}{2000}\natexlab{}.
\newblock \showarticletitle{Object-focused interaction in collaborative virtual environments}.
\newblock \bibinfo{journal}{\emph{ACM Transactions on Computer-Human Interaction}} \bibinfo{volume}{7}, \bibinfo{number}{4} (\bibinfo{date}{dec} \bibinfo{year}{2000}), \bibinfo{pages}{477–509}.
\newblock
\showISSN{1073-0516}
\urldef\tempurl%
\url{https://doi.org/10.1145/365058.365088}
\showDOI{\tempurl}


\bibitem[Kolesnichenko et~al\mbox{.}(2019)]%
        {Kolesnichenko2019}
\bibfield{author}{\bibinfo{person}{Anya Kolesnichenko}, \bibinfo{person}{Joshua McVeigh-Schultz}, {and} \bibinfo{person}{Katherine Isbister}.} \bibinfo{year}{2019}\natexlab{}.
\newblock \showarticletitle{Understanding Emerging Design Practices for Avatar Systems in the Commercial Social VR Ecology}. In \bibinfo{booktitle}{\emph{Proceedings of the 2019 on Designing Interactive Systems Conference}} \emph{(\bibinfo{series}{DIS '19})}. \bibinfo{pages}{241–252}.
\newblock
\showISBNx{9781450358507}
\urldef\tempurl%
\url{https://doi.org/10.1145/3322276.3322352}
\showDOI{\tempurl}


\bibitem[{Kurai} et~al\mbox{.}(2024)]%
        {Kurai2024-wf}
\bibfield{author}{\bibinfo{person}{{Kurai}}, \bibinfo{person}{{Hiraki}}, \bibinfo{person}{{Hiroi}}, \bibinfo{person}{{Hirao}}, \bibinfo{person}{{Perusquia-Hernandez}}, \bibinfo{person}{{Uchiyama}}, {and} \bibinfo{person}{{Kiyokawa}}.} \bibinfo{year}{2024}\natexlab{}.
\newblock \showarticletitle{Design and Implementation of Agent {APIs} for {Large-Scale} Social {VR} Platforms}. In \bibinfo{booktitle}{\emph{2024 {IEEE} Conference on Virtual Reality and {3D} User Interfaces Abstracts and Workshops ({VRW})}}, Vol.~\bibinfo{volume}{0}. \bibinfo{pages}{584--587}.
\newblock


\bibitem[Latoschik et~al\mbox{.}(2017)]%
        {Latoschik2017}
\bibfield{author}{\bibinfo{person}{Marc~Erich Latoschik}, \bibinfo{person}{Daniel Roth}, \bibinfo{person}{Dominik Gall}, \bibinfo{person}{Jascha Achenbach}, \bibinfo{person}{Thomas Waltemate}, {and} \bibinfo{person}{Mario Botsch}.} \bibinfo{year}{2017}\natexlab{}.
\newblock \showarticletitle{The effect of avatar realism in immersive social virtual realities}. In \bibinfo{booktitle}{\emph{Proceedings of the 23rd ACM Symposium on Virtual Reality Software and Technology}} \emph{(\bibinfo{series}{VRST '17})}. Article \bibinfo{articleno}{39}, \bibinfo{numpages}{10}~pages.
\newblock
\showISBNx{9781450355483}
\urldef\tempurl%
\url{https://doi.org/10.1145/3139131.3139156}
\showDOI{\tempurl}


\bibitem[{Linden Research, Inc.}(2024)]%
        {secondlife}
\bibfield{author}{\bibinfo{person}{{Linden Research, Inc.}}} \bibinfo{year}{2024}\natexlab{}.
\newblock \bibinfo{title}{Second Life}.
\newblock \bibinfo{howpublished}{\url{https://secondlife.com/}}.
\newblock
\newblock
\shownote{Accessed: 2024-6-15}.


\bibitem[{Microsoft Corporation}(2024)]%
        {minecraft}
\bibfield{author}{\bibinfo{person}{{Microsoft Corporation}}.} \bibinfo{year}{2024}\natexlab{}.
\newblock \bibinfo{title}{Minecraft}.
\newblock \bibinfo{howpublished}{\url{https://www.minecraft.net/}}.
\newblock
\newblock
\shownote{Accessed: 2024-6-15}.


\bibitem[Moustafa and Steed(2018)]%
        {Moustafa2018}
\bibfield{author}{\bibinfo{person}{Fares Moustafa} {and} \bibinfo{person}{Anthony Steed}.} \bibinfo{year}{2018}\natexlab{}.
\newblock \showarticletitle{A longitudinal study of small group interaction in social virtual reality}. In \bibinfo{booktitle}{\emph{Proceedings of the 24th ACM Symposium on Virtual Reality Software and Technology}} \emph{(\bibinfo{series}{VRST '18})}. Article \bibinfo{articleno}{22}, \bibinfo{numpages}{10}~pages.
\newblock
\showISBNx{9781450360869}
\urldef\tempurl%
\url{https://doi.org/10.1145/3281505.3281527}
\showDOI{\tempurl}


\bibitem[{NASA}(1986)]%
        {Nasa1986-vw}
\bibfield{author}{\bibinfo{person}{{NASA}}.} \bibinfo{year}{1986}\natexlab{}.
\newblock \bibinfo{title}{Nasa Task Load Index ({TLX}) v.1.0 Manual}.
\newblock
\newblock
\urldef\tempurl%
\url{http://humansystems.arc.nasa.gov/groups/TLX/downloads/TLX.pdf}
\showURL{%
\tempurl}


\bibitem[Numan et~al\mbox{.}(2023)]%
        {Numan2023-nx}
\bibfield{author}{\bibinfo{person}{Nels Numan}, \bibinfo{person}{Daniele Giunchi}, \bibinfo{person}{Benjamin Congdon}, {and} \bibinfo{person}{Anthony Steed}.} \bibinfo{year}{2023}\natexlab{}.
\newblock \showarticletitle{Ubiq-Genie: Leveraging External Frameworks for Enhanced Social VR Experiences}. In \bibinfo{booktitle}{\emph{2023 IEEE Conference on Virtual Reality and 3D User Interfaces Abstracts and Workshops (VRW)}}. \bibinfo{pages}{497--501}.
\newblock
\urldef\tempurl%
\url{https://doi.org/10.1109/VRW58643.2023.00108}
\showDOI{\tempurl}


\bibitem[Numan and Steed(2022)]%
        {Numan2022}
\bibfield{author}{\bibinfo{person}{Nels Numan} {and} \bibinfo{person}{Anthony Steed}.} \bibinfo{year}{2022}\natexlab{}.
\newblock \showarticletitle{Exploring User Behaviour in Asymmetric Collaborative Mixed Reality}. In \bibinfo{booktitle}{\emph{Proceedings of the 28th ACM Symposium on Virtual Reality Software and Technology}} \emph{(\bibinfo{series}{VRST '22})}. Article \bibinfo{articleno}{6}, \bibinfo{numpages}{11}~pages.
\newblock
\showISBNx{9781450398893}
\urldef\tempurl%
\url{https://doi.org/10.1145/3562939.3565630}
\showDOI{\tempurl}


\bibitem[Oh et~al\mbox{.}(2024)]%
        {Oh2024-gl}
\bibfield{author}{\bibinfo{person}{Jeongseok Oh}, \bibinfo{person}{Seungju Kim}, {and} \bibinfo{person}{Seungjun Kim}.} \bibinfo{year}{2024}\natexlab{}.
\newblock \showarticletitle{{LumiMood}: A Creativity Support Tool for Designing the Mood of a {3D} Scene}. In \bibinfo{booktitle}{\emph{Proceedings of the {CHI} Conference on Human Factors in Computing Systems}} \emph{(\bibinfo{series}{CHI '24})}. Article \bibinfo{articleno}{174}, \bibinfo{numpages}{21}~pages.
\newblock


\bibitem[Pan and Steed(2017)]%
        {Pan2017}
\bibfield{author}{\bibinfo{person}{Ye Pan} {and} \bibinfo{person}{Anthony Steed}.} \bibinfo{year}{2017}\natexlab{}.
\newblock \showarticletitle{The impact of self-avatars on trust and collaboration in shared virtual environments}.
\newblock \bibinfo{journal}{\emph{PLOS ONE}} \bibinfo{volume}{12}, \bibinfo{number}{12} (\bibinfo{date}{Dec.} \bibinfo{year}{2017}), \bibinfo{pages}{1--20}.
\newblock
\urldef\tempurl%
\url{https://doi.org/10.1371/journal.pone.0189078}
\showDOI{\tempurl}


\bibitem[Pearce et~al\mbox{.}(2022)]%
        {Pearce2022}
\bibfield{author}{\bibinfo{person}{Hammond Pearce}, \bibinfo{person}{Baleegh Ahmad}, \bibinfo{person}{Benjamin Tan}, \bibinfo{person}{Brendan Dolan-Gavitt}, {and} \bibinfo{person}{Ramesh Karri}.} \bibinfo{year}{2022}\natexlab{}.
\newblock \showarticletitle{Asleep at the Keyboard? Assessing the Security of GitHub Copilot's Code Contributions}. In \bibinfo{booktitle}{\emph{Proceedings of the 43rd IEEE Symposium on Security and Privacy}}. \bibinfo{pages}{754--768}.
\newblock
\urldef\tempurl%
\url{https://doi.org/10.1109/SP46214.2022.9833571}
\showDOI{\tempurl}


\bibitem[{Rec Room Inc.}(2024)]%
        {recroom}
\bibfield{author}{\bibinfo{person}{{Rec Room Inc.}}} \bibinfo{year}{2024}\natexlab{}.
\newblock \bibinfo{title}{Rec Room}.
\newblock \bibinfo{howpublished}{\url{https://recroom.com/}}.
\newblock
\newblock
\shownote{Accessed: 2024-6-15}.


\bibitem[{Roblox Corporation}(2024)]%
        {roblox}
\bibfield{author}{\bibinfo{person}{{Roblox Corporation}}.} \bibinfo{year}{2024}\natexlab{}.
\newblock \bibinfo{title}{Roblox}.
\newblock \bibinfo{howpublished}{\url{https://www.roblox.com/}}.
\newblock
\newblock
\shownote{Accessed: 2024-6-15}.


\bibitem[Shen et~al\mbox{.}(2024)]%
        {Shen2024-dq}
\bibfield{author}{\bibinfo{person}{Yulin Shen}, \bibinfo{person}{Yifei Shen}, \bibinfo{person}{Jiawen Cheng}, \bibinfo{person}{Chutian Jiang}, \bibinfo{person}{Mingming Fan}, {and} \bibinfo{person}{Zeyu Wang}.} \bibinfo{year}{2024}\natexlab{}.
\newblock \showarticletitle{Neural Canvas: Supporting Scenic Design Prototyping by Integrating {3D} Sketching and Generative {AI}}. In \bibinfo{booktitle}{\emph{Proceedings of the {CHI} Conference on Human Factors in Computing Systems}} \emph{(\bibinfo{series}{CHI '24})}. Article \bibinfo{articleno}{1056}, \bibinfo{numpages}{18}~pages.
\newblock


\bibitem[{Solirax}(2024)]%
        {neos}
\bibfield{author}{\bibinfo{person}{{Solirax}}.} \bibinfo{year}{2024}\natexlab{}.
\newblock \bibinfo{title}{Neos}.
\newblock \bibinfo{howpublished}{\url{https://neos.com/}}.
\newblock
\newblock
\shownote{Accessed: 2024-6-15}.


\bibitem[Steed et~al\mbox{.}(2022)]%
        {Steed2022}
\bibfield{author}{\bibinfo{person}{Anthony Steed}, \bibinfo{person}{Lisa Izzouzi}, \bibinfo{person}{Klara Brandstätter}, \bibinfo{person}{Sebastian Friston}, \bibinfo{person}{Ben Congdon}, \bibinfo{person}{Otto Olkkonen}, \bibinfo{person}{Daniele Giunchi}, \bibinfo{person}{Nels Numan}, {and} \bibinfo{person}{David Swapp}.} \bibinfo{year}{2022}\natexlab{}.
\newblock \showarticletitle{Ubiq-exp: A toolkit to build and run remote and distributed mixed reality experiments}.
\newblock \bibinfo{journal}{\emph{Frontiers in Virtual Reality}}  \bibinfo{volume}{3} (\bibinfo{year}{2022}).
\newblock
\showISSN{2673-4192}
\urldef\tempurl%
\url{https://doi.org/10.3389/frvir.2022.912078}
\showDOI{\tempurl}


\bibitem[{Tabnine Ltd.}(2024)]%
        {tabnine}
\bibfield{author}{\bibinfo{person}{{Tabnine Ltd.}}} \bibinfo{year}{2024}\natexlab{}.
\newblock \bibinfo{title}{Tabnine}.
\newblock \bibinfo{howpublished}{\url{https://www.tabnine.com/}}.
\newblock
\newblock
\shownote{Accessed: 2024-6-15}.


\bibitem[{Unity Technologies}(2024)]%
        {unity}
\bibfield{author}{\bibinfo{person}{{Unity Technologies}}.} \bibinfo{year}{2024}\natexlab{}.
\newblock \bibinfo{title}{Unity}.
\newblock \bibinfo{howpublished}{\url{https://unity.com/}}.
\newblock
\newblock
\shownote{Accessed: 2024-6-15}.


\bibitem[{VRChat Inc.}(2024)]%
        {vrchat}
\bibfield{author}{\bibinfo{person}{{VRChat Inc.}}} \bibinfo{year}{2024}\natexlab{}.
\newblock \bibinfo{title}{{VRChat}}.
\newblock \bibinfo{howpublished}{\url{https://hello.vrchat.com/}}.
\newblock
\newblock
\shownote{Accessed: 2024-6-15}.


\bibitem[Yan et~al\mbox{.}(2023)]%
        {Yan2023-sz}
\bibfield{author}{\bibinfo{person}{Dapeng Yan}, \bibinfo{person}{Zhipeng Gao}, {and} \bibinfo{person}{Zhiming Liu}.} \bibinfo{year}{2023}\natexlab{}.
\newblock \showarticletitle{A Closer Look at Different Difficulty Levels Code Generation Abilities of {ChatGPT}}. In \bibinfo{booktitle}{\emph{Proceedings of the 38th {IEEE/ACM} International Conference on Automated Software Engineering ({ASE})}}. \bibinfo{pages}{1887--1898}.
\newblock


\bibitem[Yee et~al\mbox{.}(2007)]%
        {Yee2007}
\bibfield{author}{\bibinfo{person}{Nick Yee}, \bibinfo{person}{Jeremy~N. Bailenson}, \bibinfo{person}{Mark Urbanek}, \bibinfo{person}{Francis Chang}, {and} \bibinfo{person}{Dan Merget}.} \bibinfo{year}{2007}\natexlab{}.
\newblock \showarticletitle{The Unbearable Likeness of Being Digital: The Persistence of Nonverbal Social Norms in Online Virtual Environments}.
\newblock \bibinfo{journal}{\emph{CyberPsychology \& Behavior}} \bibinfo{volume}{10}, \bibinfo{number}{1} (\bibinfo{year}{2007}), \bibinfo{pages}{115--121}.
\newblock
\urldef\tempurl%
\url{https://doi.org/10.1089/cpb.2006.9984}
\showDOI{\tempurl}


\bibitem[{Yellow Dog Man Studios}(2024)]%
        {resonite}
\bibfield{author}{\bibinfo{person}{{Yellow Dog Man Studios}}.} \bibinfo{year}{2024}\natexlab{}.
\newblock \bibinfo{title}{Resonite}.
\newblock \bibinfo{howpublished}{\url{https://resonite.com/}}.
\newblock
\newblock
\shownote{Accessed: 2024-6-15}.


\end{thebibliography}


\end{document}